\documentclass[twoside]{dis08}
\usepackage[latin1]{inputenc}
\usepackage[dvips]{graphicx,epsfig,color}
\usepackage{wrapfig,rotating}
\usepackage{amssymb,amsmath,array}

\begin{document}
\title{Spin Density Matrix Elements from diffractive $\phi$ meson 
production at HERMES
}

\author{W. Augustyniak$^1$, A. Borissov$^{2}$\thanks{For A.B. this work is supported in part by the Heisenberg-Landau program. } \; and S. Manayenkov$^3$, \\
on behalf of the HERMES Collaboration
%
%
\vspace{.3cm}\\
%
1- 00-681 Warsaw,Hoza 69,Poland  - The Andrzej Soltan Institute for
Nuclear Studies\\
%
2- DESY, D-22607 Hamburg, Germany \\
3- Gatchina, Leningrad region, 188300, Russia -
Petersburg Nuclear Physics Institute \\
}

\maketitle

\begin{abstract}
Exclusive production of $\phi$ mesons 
on hydrogen and deuterium targets is studied in the HERMES
kinematic region  $1 < Q^{2} < 7$ GeV$^2$ and $3.0 < W < 6.3$ GeV.
Spin density matrix elements 
and their $Q^{2}$ and $t$ dependences are presented.
These data are consistent with $s$-channel helicity conservation in 
exclusive $\phi$ meson production.
No statistically significant evidence 
for the contribution of unnatural-parity-exchange amplitudes is found.
\end{abstract}

\section{Introduction}
Electroproduction of neutral vector mesons from nucleons is
  described by the interaction with the nucleon of
  $ q \bar q$  pair created by virtual photon.
The reaction  $e+N \rightarrow  e^{\prime}+\phi +N'$  in 
the single-photon approximation is equivalent to 
$\gamma^{*} + N \rightarrow \phi +N'$, in which the $\gamma^{*}$ has 
negative squared four-momentum transfer
$Q^{2}$ and
polarization parameter $\epsilon$,
the ratio of fluxes of longitudinal and transverse virtual photons. 
The spin transfer from the virtual photon $\gamma^{*}$ 
to the vector meson V is commonly described~\cite{wolf,mdiehl} in terms of 
Spin Density Matrix Elements (SDMEs).
The experimentally determined  set of 23 SDMEs for the $\phi$ meson 
is presented here
in the Schilling-Wolf \cite{wolf} representation.
From measured values of the SDMEs one may examine
$s$-Channel Helicity Conservation (SCHC) in the transition 
$\gamma^* \rightarrow$V. The fractional 
contribution of Unnatural-Parity-Exchange (UPE) amplitudes of the process 
$\gamma^{*}+N \rightarrow V +N$
in comparison with 
Natural-Parity-Exchange (NPE) amplitudes can also be derived from 
SDMEs measurements. 
Natural-parity exchange  indicates that the interaction is mediated by a 
particle of ``natural''
parity ($ J^{P} =0^{+},1^{-},2^{+}, ...$),
while  UPE amplitudes describe the exchange of a particle of  
``unnatural'' parity 
($J^{P} =0^{-},1^{+}, ... $).
Since the $\phi$ meson consists mainly of strange quarks with a very
small admixture of other quark flavors,  
it is expected that quark-exchange with the nucleon is suppressed
and two- (or more) gluon exchange dominates. 
For $\rho^{0}$ meson production at HERMES kinematics,
both two-gluon-exchange and quark-exchange contributions of the same order of 
magnitude~\cite{golos3,abb} are needed to describe the data.
Hence a comparison of SDMEs for $\phi$ and $\rho ^0$ mesons allows a test of our
understanding of these mechanisms.
Further information about these final-state interactions can be obtained 
by measuring the 
phase difference between amplitudes.

\section{Selection of exclusive $\phi$ mesons} 

\begin{wrapfigure}{r}{0.6\columnwidth}
\centerline{\includegraphics[width=0.56 \columnwidth]{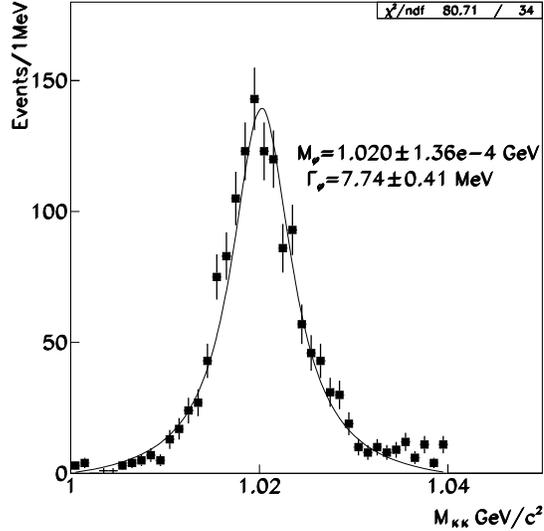}}
\caption{Two-kaon invariant mass distribution fitted with a Breit-Wigner function. 
The fit indicates a  negligible non-resonant contribution. }
\label{Fig:DELE}
\end{wrapfigure}

The experiment was performed with longitudinally polarized electron and positron beams at 
an energy of 27.5 GeV 
using unpolarized hydrogen or deuterium gas targets.
The decay products of the $\phi$ mesons 
($\phi \rightarrow K^{+}K^{-}$, branching ratio $\approx 49\%$) 
were detected in the HERMES spectrometer \cite{Det1}.
The $\phi$ mesons were selected by requiring
$ 0.99<M_{KK}<1.04$ GeV for the  invariant mass of the two hadrons
(see Fig.~\ref{Fig:DELE}).  
In addition, the identification of the decay particles was required. 
For the data set obtained in 1996-1997,
the absence of a threshold pion signal from the Cherenkov detector was required, 
while for the data from 2000-2002  information from the RICH \cite{Rich} detector
was used.
As can be seen from  Fig.~\ref{Fig:DELE} 
the non-resonant contribution under the $\phi$ meson invariant mass peak is negligible. 
Exclusive events were selected by 
the requirement:
$\Delta E = (M_X^2-M^2)/(2M) \leq 0.6$ GeV, 
where $M$ ($M_X$) is the
mass of the nucleon (missing mass). 
Diffractive events were selected by the constraint 
$-t' < 0.4$ GeV$^2$.
Here, $t'=t-t_0$, with $t$ being the squared four-momentum  transfer
from virtual photon to vector meson and $-t_0$ 
the smallest kinematically allowed value of $-t$ at 
fixed $Q^2$ and energy in the $\gamma^* N$ center-of-mass 
 system ($W$). 
The fraction of semi-inclusive deep-inelastic scattering events in the chosen region was 
determined to be 1.6\%.
\section{Spin Density Matrix Elements}

\subsection{Formalism}
The matrix $r^{\alpha}_{\lambda_{V}\lambda^{'}_{V}}$  is given by 
$   r^{\alpha}_{\lambda_{V}\lambda^{'}_{V}}=\frac{1}{2 \mathcal{N} }
\sum_{\lambda_{\gamma}\lambda^{'}_{\gamma}\lambda_{N}\lambda^{'}_{N}}
F_{\lambda_{V} \lambda^{'}_{N}\lambda_{\gamma}\lambda_{N}}
\Sigma^{\alpha}_{\lambda_{\gamma} \lambda^{'}_{\gamma}}
F^{*}_{\lambda^{'}_{V} \lambda^{'}_{N}\lambda^{'}_{\gamma} \lambda_{N}},
$
where 
$F_{\lambda_{V} \lambda^{'}_{N}\lambda_{\gamma}\lambda_{N}}$ is the amplitude of 
the process
$\gamma ^{*} (\lambda_{\gamma})+N(\lambda_{N}) \rightarrow V(\lambda_{V})+N^{'}(\lambda^{'}_{N})$ 
(symbols 
in parenthesis denote  particle helicities) and $\mathcal{N}$ is the normalization factor.
The amplitude $F$ may be decomposed into the sum of a NPE 
amplitude
$T_{\lambda_{V} \lambda^{'}_{N}\lambda_{\gamma}\lambda_{N}}$ and a UPE amplitude 
$U_{\lambda_{V} \lambda^{'}_{N}\lambda_{\gamma}\lambda_{N}}$. Since the contribution 
to the SDMEs 
from the NPE 
amplitudes with nucleon spin flip are suppressed by a factor of about $-t'/(4M^2)$,
only the NPE 
amplitudes with
$\lambda^{'}_{N}=\lambda_{N}$ will be discussed here and the shorthand notation 
$T_{\lambda_{V}\lambda_{\gamma}}=T_{\lambda_{V}\frac{1}{2}\lambda_{\gamma}\frac{1}{2}}$ 
will be used below. 
The nine Hermitian matrices $\Sigma^{\alpha}_{\lambda_{\gamma} \lambda^{'}_{\gamma}}$ are defined in 
\cite{wolf}.  
The  upper index $\alpha=0 $ corresponds to the unpolarized transverse photon, 
$\alpha=1,2$ to the two directions
of linear polarization, $\alpha=3$ - circular polarization, $\alpha=4$ corresponds to the 
longitudinal 
photon and $\alpha=5-8$  are attributable to the interference of longitudinal and transverse 
photons.
Without using different lepton beam energies contributions of the transverse and 
longitudinal 
photons cannot be disentangled, 
hence the 
matrices $r^0_{ij}$ and $r^4_{ij}$ cannot be measured 
separately; only their combination $r^{04}_{ij} =r^0_{ij}+\epsilon r^4_{ij}$ is determined.
For this case the normalization 
factor $\mathcal{N}$ can be found from the condition 
${\rm tr} \{r^{04}\}=r^{04}_{00}+r^{04}_{11}+r^{04}_{-1-1}=1$. 
Eight matrix elements of the matrices $r^3_{ij}$, $r^7_{ij}$, $r^8_{ij}$, refered to as ``polarized'' SDMEs, 
measured for the first time by HERMES, are presented in Fig.~\ref{Fig:sdme2} (yellow background). 
The other 15 matrix elements, called ``unpolarized,''
do not require a polarized beam.

The 3-dimensional angular distribution of the scattered lepton and  the decay products
is described by the following angles (see detailed definitions in \cite{joos}): $\Phi$
is the angle between the $\phi$ meson  production and lepton 
scattering planes in the $\gamma^{*}$p center-of-mass system.
The polar  $\varphi$ and azimuthal  $\Theta$
angles of the decay $K^{+}$ are presented in the $\phi$-meson rest frame.

The extracted SDMEs will be presented based on a hierarchy of NPE helicity
amplitudes:
$ |T_{00}| \sim |T_{11}| \gg |T_{01}| > |T_{10}| \sim |T_{1-1}|$,
established for the first time in Ref.~\cite{divanov} for $\rho ^0$ production. 
This hierarchy is valid
at asymptotically high value of $Q^2$, and was experimentally confirmed for 
exclusive $\rho^0$ production at HERMES kinematics~\cite{abb}.
The SDMEs are categorized into five classes according to this 
hierarchy. 
The two classes A and B  describe only SCHC transitions. Classes from C  to E contain
also spin flip transitions.   
Class A comprises SDMEs with dominant contributions proportional to $|T_{00}|^2$ or $|T_{11}|^2$ - the 
helicity-con\-ser\-ving amplitudes 
describing the transitions $\gamma^*_L \to V_L$ and $\gamma^*_T \to V_T$.
Class B  SDMEs  correspond to the interference of   $T_{00}$ and $T_{11}$
amplitudes. The main terms  for the unpolarized (polarized) SDMEs are proportional to the real
(imaginary) part of $T_{00}T_{11}^*$. In fact, as a general rule for the classes B - E, the dominant
contribution of the unpolarized (polarized) SDMEs is proportional to the real (imaginary) part of
the product of two amplitudes.
Class C contains SDMEs with dominant terms that are products 
of the $s$-channel helicity non-conserving  amplitude $T_{01}$
(corresponding to the $\gamma^*_{T} \to V_{L}$ transition), 
 and $T_{00}^*$ or  $T_{11}^*$ 
(for $r^1_{00}$ the $T_{01}$ contribution is actually quadratic).
Classes D and E are composed of  SDMEs in which the main terms contain a
product of the small helicity-flip amplitudes
$T_{10}$ ($\gamma^*_{L} \to V_{T}$)
and $T_{-11}$  ($\gamma^*_{T} \to V_{-T}$), respectively, multiplied by $T_{11}^*$.

\subsection{Extracted SDMEs}
The SDMEs are obtained 
 by minimizing the difference between the
3-dimensional $(\cos\Theta,\phi,\Phi)$  matrix of the
data and the analogous matrix from fully reconstructed simulated
events.   
An $8 \times 8 \times 8$
binning was used for the  variables $\cos\Theta$, $\phi$, $\Phi$.
The simulated events were generated with uniform angular distributions
and  re-weighted in an iterative procedure with the angular distribution
$\mathcal{W} (\cos\Theta,\phi,\Phi,r^{\alpha}_{ij} )$~\cite{wolf}, where the spin density matrix
elements $r^{\alpha}_{\lambda_V \lambda^{'}_V}$ were treated as free parameters. 
The best fit parameters were determined using a binned maximum log-likelihood method.
The minimization itself and the error
calculation were performed by MINUIT.
\begin{figure}[bottom]
\begin{center}
\includegraphics[width=11.5cm,height=10.5cm]{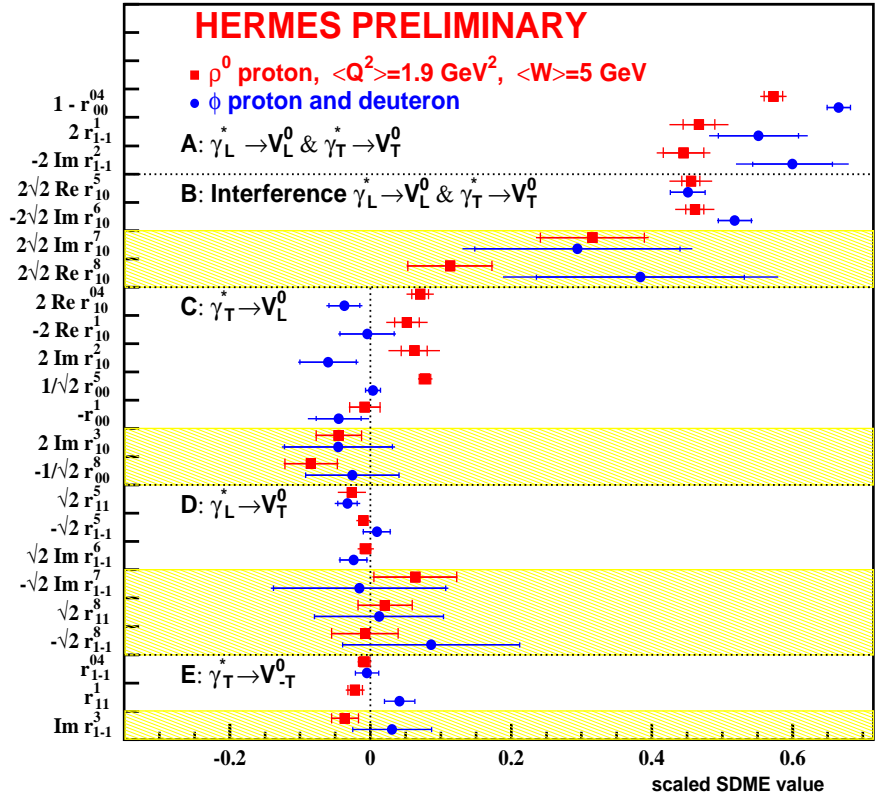}
\caption{ The 23 SDME's extracted for $\rho^{0}$ production on the proton (squares)
and for $\phi$ meson production on proton and deuteron (circles). 
The inner error bars represent 
the statistical uncertainties, while the outer ones indicate the statistical and 
systematic uncertainties added in quadrature. 
The unshaded (shaded) areas indicate unpolarized (polarized) SDME's.
For easier interpretation the set of SDMEs was divided in five classes (see text). 
}
\label{Fig:sdme2}
\end{center}
\end{figure}

The extracted SDMEs for the kinematic region
 $1 < Q^{2} < 7$ GeV$^2$, $3< W < 6.3$ GeV, 
and $0 < -t^{\prime} < 0.4$ GeV$^2$, 
are presented for $\rho^0$ and $\phi$ meson data in Fig.~\ref{Fig:sdme2}.
The average kinematic values are $\langle Q^{2} \rangle= 1.9$ GeV$^2$, $\langle W \rangle=4.8$ GeV
and $\langle -t^{\prime} \rangle = 0.13$ GeV$^2$.
The experimental uncertainties are
larger for the eight polarized SDMEs due
to the imperfect lepton beam polarization (0.53), and the small kinematic factor $\sqrt{1-\epsilon}$
($\langle \epsilon \rangle= 0.8$) by which the polarized SDMEs are multiplied.

In Fig.~\ref{Fig:sdme2} the SDMEs are shown multiplied by certain factors to make the coefficients
of the dominant amplitude products equal to unity.
The elements of class A 
 are presented in the figure in such a way that their 
main terms are proportional to $|T_{11}|^2$, in particular $1 - r^{04}_{00}$ is chosen. 
The  SDMEs of class A  are similar for $\rho ^0$ and $\phi$. The elements of class 
B are also close to each other for both vector mesons. 
For $\phi$ mesons the values of the elements
of classes C, D and E fluctuate near zero
supporting the validity of SCHC. 
For  $\rho^0$  mesons  the elements of classes C, D, E
with significantly non-zero values indicate that there exists
also a production mechanism  which
does not conserve $s$-channel helicity. We note, that small violation of SCHC was 
observed for $\phi$ production  
by the H1 Collaboration \cite{H1phi}.

\begin{figure}    
\includegraphics[width=12.cm,height=9.5cm]{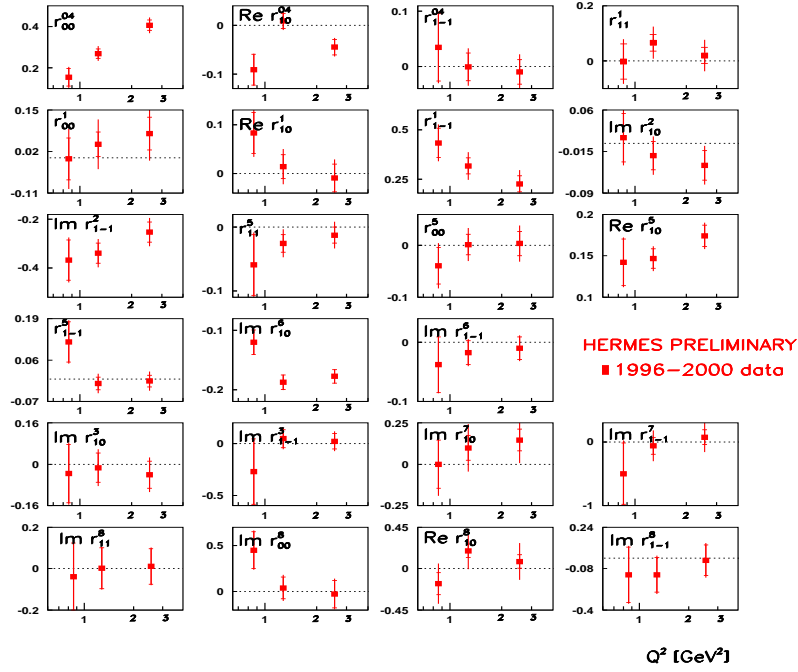}
\vspace*{-0.5cm}
\includegraphics[width=12.cm,height=9.5cm]{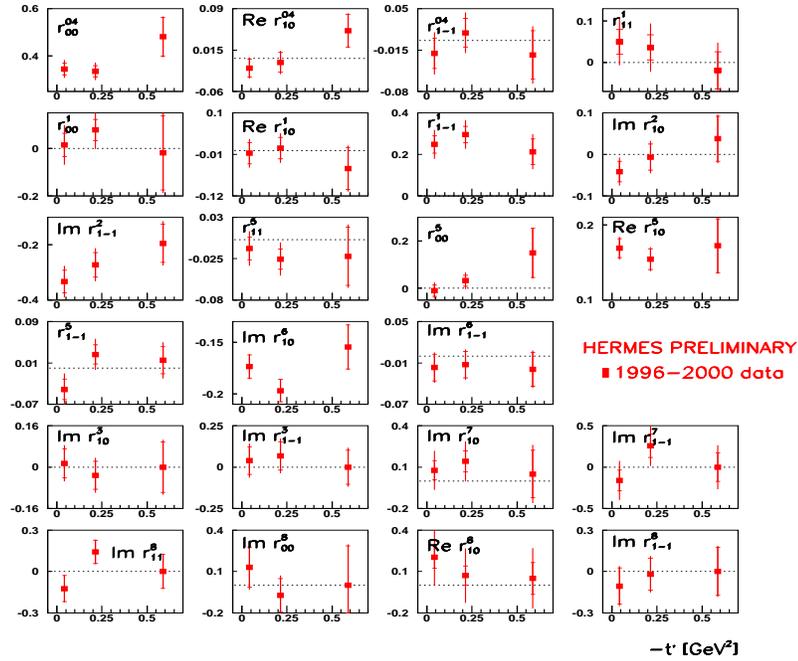}
\caption{The kinematic dependence  of $\phi$ SDMEs on $Q^{2}$ and $t^{\prime}$.
Proton and deuteron data are combined. The inner error bars represent 
the statistical uncertainties, while the outer ones indicate the statistical and 
systematic uncertainties added in quadrature.
}
\label{Fig:Q2dep}
\end{figure}
The dependences of $\phi$ meson SDMEs on $Q^{2}$ and $t^{\prime}$ are shown in Fig.~\ref{Fig:Q2dep}.
For the presentation of the $Q^2$ ($t'$) dependence, an additional bin of $0.5 < Q^2 < 1$ 
GeV$^2$ ( $0.4< -t'< 1$ GeV$^2$) is included.
As seen also in Fig.~\ref{Fig:sdme2}, the SDMEs  
from  classes A and B are non-zero (in agreement with a GPD model calculation~\cite{golos3}). 
One can see some dependence on $Q^{2}$ for these elements, however a $t'$ dependence is not observed.
We note that the SCHC relations  $r^{1}_{1-1}=-{\rm Im}\{r^{2}_{1-1}\}$,
${\rm Re}\{r^{5}_{10}\}=-{\rm Im}\{r^{6}_{10}\}$, ${\rm Re}\{r^{8}_{10}\}={\rm Im}\{r^{7}_{10}\}$ are
fulfilled in every kinematic bin in $Q^2$ and $t'$.
The elements from classes  C to E,
related  to  spin-flip transitions, should mainly depend on $t^{\prime}$; the present experimental
uncertainty is insufficient to observe any dependence.

\subsection{ Phase Difference of $T_{11}$ and $T_{00}$}
The interference between the dominant amplitudes $T_{11}$ and $T_{00}$ 
depends on their phase  difference $\delta$ which can be evaluated as follows:
${\rm tan}\; \delta = 
[{\rm Im} \{r^7_{10}\} +{\rm Re} \{r^{8}_{10} \} ]/
[{\rm Re} \{r^5_{10} \} -{\rm Im}\{r^{6}_{10}\} ].$
This results in
$\delta= 33.0 \pm 7.4$~deg.\ for the combined proton and deuteron $\phi$ meson data.
For the first time,  the sign of $\delta$ is determined to be positive.
We note that in the GPD-based calculations of Ref.~\cite{golos3}, the 
value of $\delta$  is found to be  $\sim 3$~degrees. This appears to be
inconsistent with the above result.

\subsection{Unnatural-Parity Exchange}
Without assuming SCHC, the hypothesis of natural parity exchange in the t channel
alone leads to the following relations: 
$U_{1}=1-r^{04}_{00} -2r^{04}_{1-1} -2r^{1}_{11} -2r^{1}_{1-1} =0$,
$U_{2}=r^{5}_{11}+r^{5}_{1-1}=0$, and $U_{3}=r^{8}_{11}+r^{8}_{1-1}=0$.\\
\begin{wrapfigure}{r}{0.55\columnwidth}
\centerline{\includegraphics[width=0.65\columnwidth]{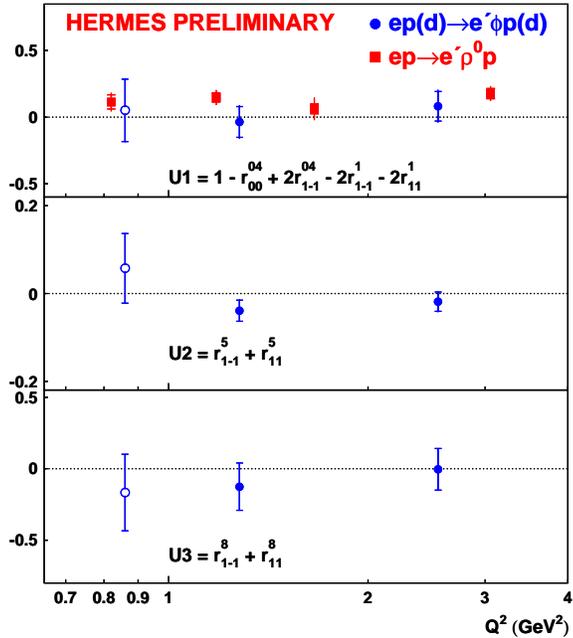}}
\caption{ $Q^{2}$ dependence of the linear combinations of
SDMEs $U_1$, $U_2$, $U_3$ for $\phi$ and  $\rho^0$ (only on the top panel). 
 Values integrated over the range $1 < Q^2 < 7$ GeV$^2$  
are presented in the first bin $Q^2 < 1$ GeV$^2$ by open symbols.
Only statistical errors are presented.
}
\label{Fig:npe}
\end{wrapfigure}
As presented in Fig.~\ref{Fig:npe},
all those relations are approximately fulfilled, 
indicating that UPE amplitudes do not contribute significantly to exclusive $\phi$ meson production.
A non-zero signal of UPE is evidence for the existence of quark-antiquark exchange, 
which is observed for exclusive $\rho^0$ (see Fig.~\ref{Fig:npe}).

\begin{footnotesize}



%

\end{footnotesize}


\end{document}